\documentclass {jaa}
\usepackage[english]{babel} 
\usepackage{graphics}
\usepackage{graphicx}
\usepackage{color}
\usepackage{amsmath}
\usepackage{psfrag}
\usepackage{epsfig}
\usepackage[round]{natbib}

\newcommand{\rsun}{R$_\odot$}

\begin{document}
\title[The Ooty Wide Field Array ]{The Ooty Wide Field Array}
\author[Subrahmanya, Manoharan \& Chengalur ] {C. R. Subrahmanya$^{1}$\thanks{Email:crs@rri.res.in}, P. K. Manoharan$^{2}$\thanks{Email:mano@ncra.tifr.res.in}, Jayaram. N. Chengalur$^{3}$\thanks{Email:chengalur@ncra.tifr.res.in}\\ 
  $^{1}$ Raman Research Institute, C. V. Raman Avenue, Sadashivnagar,Bengaluru 560080,India. \\
  $^{2}$ Radio Astronomy Centre, NCRA-TIFR, P. O. Box 8, Ooty 643001, India. \\
    $^{3}$ NCRA-TIFR, Pune University Campus, Ganeshkhind, Pune 411007,India. \\
}
\date {}
\maketitle

\begin{abstract}

We describe here an ongoing upgrade to the legacy Ooty Radio Telescope (ORT).
The ORT is a cylindrical parabolic cylinder 530mx30m in size operating at 
a frequency of 326.5 (or $z \sim 3.35$ for the HI 21cm line). The telescope
has been constructed on a north-south hill slope whose gradient is equal to the 
latitude of the hill, making it effectively equitorially mounted. The 
feed consists of an array of 1056 dipoles. The key feature of this upgrade
is the digitisation and cross-correlation of the signals of every set of
4-dipoles. This converts the ORT into a 264 element interferometer with a
field of view of $ 2^{\circ} \times 27.4^{\circ} \cos(\delta)$. This upgraded
instrument is called the Ooty Wide Field Array (OWFA). This paper briefly
describes the salient features of the upgrade, as well as its main science
drivers. There are three main science drivers viz. (1) Observations of the
large scale distribution of HI in the post-reionisation era (2) studies of the 
propagation of plasma irregularities through the inner heliosphere and 
(3) blind surveys for transient sources. More details on the upgrade,
as well as on the expected science uses can be found in other papers
in this special issue.

\end{abstract}


\section{Introduction}    

\begin{figure}[b!]
\centerline{\includegraphics[width=7cm]{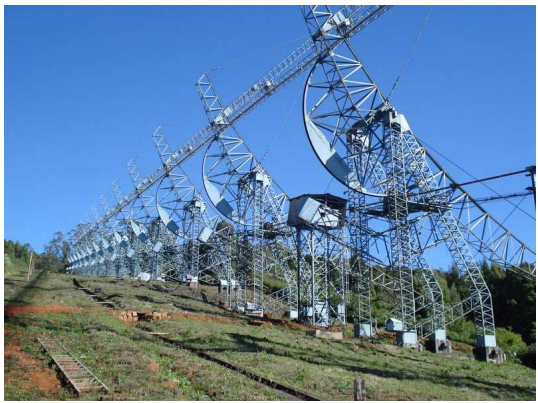}}
\label{fig:ORT}
\caption{ The Ooty Radio Telescope is an offset cylindrical paraboloid,
  530mx30m in size. It is equitorially mounted. The Ooty Wide Field
  Array~(OWFA) is an upgrade to the ORT, which digitizes the signals from
  every 4 dipoles along the feed line. These signals are then cross-correlated,
  leading to a very wide field of view interferometer. See the text for more
  details.
  }
\end{figure}

This article describes the Ooty Wide Field Array (OWFA) which is an
ongoing upgrade of  the  Ooty Radio Telescope (ORT). The ORT \citep{swarup71}
is an offset parabolic cylinder 530mx30m in size commissioned in 1970. The 
telescope is mechanically steerable about the long axis of the cylinder. 
This axis is mounted along a hill whose slope is the same as its latitude, 
effectively making the telescope equitorially mounted. The sky can hence
be tracked by rotation about the cylinder axis. The telescope has a line 
focus, along which is placed an array of 1056 dipoles. Each dipole is 
followed by a Low Noise Amplifier (LNA) as well as a 4-bit phase 
shifter\citep{selvanayagam93}. The dipoles are grouped into 22 ``modules'' 
with each module consisting of 48 dipoles. In the legacy system the phased 
output of all modules are combined with appropriate phase and delay correction 
to produce up to 12 equispaced beams covering 36$^{'}$ in the north-south
direction. The delays and phases can also be adjusted to change the
North-South position of the beam centroid. The telescope pointing is hence
a combination of mechanical steering (along the east west direction) and
electrical steering (along the north-south direction). The legacy telescope
had an operational frequency centred at  326.5~MHz and a usable bandwidth
of $\sim 4$~MHz.

Compared to the legacy ORT the OWFA will have both a larger instantaneous bandwidth
as well as a significantly larger field of view. The main science drivers
behind this upgrade are (1) detection of the HI 21cm emission from the large
scale structure at $z \sim 3.3$, (2) Monitoring of the weather in the inner
solar heliosphere (via high cadence observations of a dense grid of 
scintillating extra-galactic radio sources) and (3) searches for transient 
sources. All three of these science drivers are described in more detail in 
other articles in this issue. Here we give an overview of OWFA itself and a 
summary of its effectiveness in achieving these main science goals.

\section{The OWFA Instrument}

The OWFA (see Table 1) will operate as an interferometer, where each element
of the interferometer
consists of the phased output of 4 contiguous dipoles from the feed array.
This corresponds to a length of 1.92m., so the geometric area corresponding
to each element is 1.92x30m. The effective system  temperature is 
150~K \citep{selvanayagam93}.  The process of conditioning and digitising 
this signal is described briefly below, the interested reader is referred 
to \citet{subrahmanya16} for more details. 

The phased output of the 4 dipole system is amplified and transported
from the focal line to a set of ``pillars'' located below the reflecting
surface by co-axial cables. There is one pillar for each module, catering
to 12 elements or a total of 48 dipoles. The set of 12 signals received
at each pullar are amplified and then fed into a 12 channel ADC. SAW filters 
with a centre frequency of 327.5 MHz are used for image rejection, and 
the signals are bandpass sampled to give an effective bandwidth of 38~MHz.
The sampling clock in each of the pillars is locked to the central time
and frequency standard of the observatory. A 1pps signal is also distributed
to all of the pillars to allow for synchronisation. The digitised signals 
from the ADC are packetized using an onboard FPGA. These packets are
transported to the central receiver room via optical fibre, using the
Xlinx aurora protocol. 

     In the central receiver room the data processing is done using
a Networked Signal Processing System (NSPS) \citep{prasad11} which
consists of a set of custom FPGA cards and an HPC. The cards rearrange 
the data into sets of time slices, with each timeslice containing 350~ms 
worth of data data from all 264 elements re-framed as standard Ethernet
UDP packets. These data are sent out using 1 GB Ethernet links to an 8
node HPC, with each node receiving one timeslice. Each HPC node consists 
of a dual socket Haswell sever, along with an add on Intel Xeon-Phi 3510 card. 
This design allows for embarrassingly parallel data processing, with
an 800 spectral channel, 264 element software FX correlator on
each node with the F engine being on the mother board and the X engine
on the Xeon-Phi cards.

\begin{table}
\begin{center}
\caption{System parameters for OWFA}
\vspace{.2in}
\label{tab:array}
\begin{tabular}[scale=.3]{|l|c|}
\hline \hline Parameter & Value\\ 
\hline No. of elements  & 264 \\ 
\hline System Temp.     & $150\,{\rm K}$ \\
\hline Aperture area    & $ 30 \,{\rm m} \times 1.92 \,{\rm m}$ \\
\hline Field of View    & $ 1.75^{\circ} \times 27.4^{\circ} \cos(\delta)$\\ 
\hline Shortest spacing & $1.9 \,{\rm m}$ \\
\hline Longest spacing  & $505.0 \,{\rm  m}$\\
\hline Total Bandwidth  & $38\,{\rm MHz}$ \\
\hline Spectral Resolution& $48\,{\rm kHz}$\\
\hline
\hline
\end{tabular}
\end{center}
\end{table}

There was a staged path to the realisation of OWFA. One of the first stages
in the path was a precursor system (the so called ``Phase~I'' system), which
consisted of 40 elements, each corresponding to the combination of the
signals from 24 dipoles (or half a module) and had a bandwidth of
$\sim 19$~MHz. The Phase~I system did not include the two extreme modules
of the ORT, and has been described in detail by \citet{prasad11a}.  Some of
the papers in this  issue also refer to this system. The Phase~I system
is no longer in active use.  Since in the initial phases of observation
there may be some advantages in testing algorithms using a  simpler
system where each element has a  higher sensitivity and a smaller field
of view, it is planned either that the Phase~I system will be restored or
that a similar functionality will be made available using the OWFA hardware.

\begin{figure}
\centerline{\includegraphics[width=7cm]{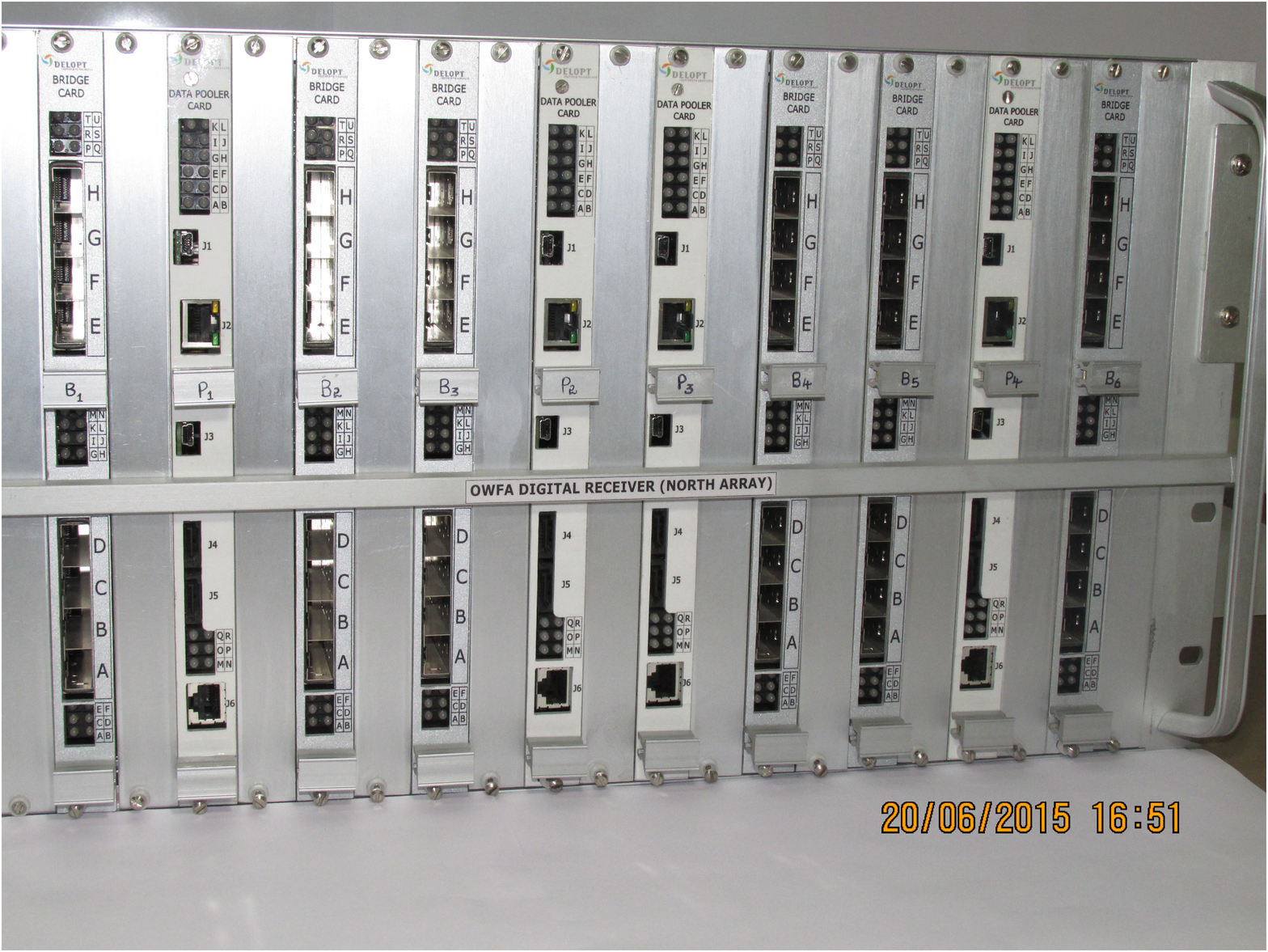}}
\label{fig:OWFARack}
\caption{ Custom FPGA rack to handle the data from the north half of OWFA.
  There is a similar rack for the south half of the telescope. These cards
  re-organize the packetized data from the 264 elements into 8 time-slices,
  with each time-slice containing the data from all 264 elements. The time
  slices are transferred to 8 different HPC nodes via ethernet. This allows
  the HPC nodes to independently carry out the 264 element FX correlation
  for each time-slice.
  }
\end{figure}

The current status of the instrument is as follows. All of the RF amplifiers
have been installed in the field and have been in regular use for some time.
In fact (as detailed in \cite{subrahmanya16}) these amplifiers now form
a critical element in the path of the legacy system, and as such have 
been rigorously tested in the field. The digital systems (both in
the field as well as in the central building) have been installed and 
interconnected by optical fibre links. The distribution of the sampling
clock and the 1pps synchronising signals to the different pillars has been
completed and tested. The data flow from the ADCs in the field to the
HPC has also been tested. The key elements of the software correlator have
been written, and it is expected that the instrument would be taking sky
data by the mid 2017.

\section{Key Science Goals of OWFA}

\subsection{HI at z ~ 3.3}

The HI 21cm line is emerging as an important probe of both cosmological
parameters as well as structure formation over a wide redshift range
(see \citet{pritchard12} for a review). Importantly, the HI 21cm line is the
only available probe in during the ``cosmic dark ages'', i.e. before the 
formation of the first ionising agents, and remains one of the most 
important probes through the ``cosmic dawn'' era (when the first ionising 
stars and blackholes form) and the epoch of reionisation. In the 
post-reionisation era neutral hydrogen is largely confined to dense collapsed
objects but its large scale distribution remains an important probe
of both structure formation and the evolution of cosmic parameters
(see e.g. \cite{bharadwaj01,bharadwaj01a, bharadwaj09,chang10,bagla10}). The
cosmology enabled using the HI~21cm line (along of course, with other
science drivers) has lead to a reawakening of interest in low frequency
radio astronomy, and a number of experiments are being planned in order
to probe these different cosmic epochs. Experiments planned for the
post-reionisation era include CHIME (\citet{Bandura2014}), 
BAOBAB (\citet{Pober2013})  and Tianlai CRT (\citet{Chen2011}).
The wide field of view, excellent sampling of the short baselines (where
the HI signal is expected to be strong \citep{ali14}) as well as good 
sensitivity make OWFA also well suited for observations of the 
post-reionisation HI signal. The highly redundant configuration of the 
OWFA interferometer also allows for efficient and high cadence redundant 
calibration \citep{marthi14}, which is important in reducing systematics 
arising from foregrounds. Detailed calculations of the expected signal 
show that OWFA sensitivity is sufficient to make a $\sim 5\sigma$ detection
of the amplitude of the power spectrum A$_{\rm HI}$  of the large scale
distribution of HI emission at $z\sim 3.3$ in integration times of 
$\sim 150$hr. Longer integrations ($\sim 1000$~hrs \citep{sarkar16,ghelot16}), 
should have the sensitivity to measure the power spectrum at angular 
scales between $11^{'}$ and $3^{o}$ (or wavenumbers from $\sim 0.02$
to $0.5$ Mpc$^{-1}$).  Two important astro-physical
parameters that are constrained by these observations are the amplitude
of the power spectrum A$_{\rm HI}$ (which in turn depends on the cosmic
density of neutral hydrogen ($\Omega_{\rm HI}$), the neutral fraction
(x$_{\rm HI}$), and the bias parameter (b$_{\rm HI}$)) and the redshift
distortion parameter $\beta$ \citep{bharadwaj15}. As mentioned above,
these measurements in turn also lead to constraints on the cosmological
parameters \citep{bharadwaj09}. The expected HI 21cm signal is however
many orders of magnitude fainter than the radio emission from
other foreground sources such as the diffuse galactic synchrotron emission
and discrete  extra-galactic radio sources. A detailed study of these
foregrounds and their effect on the expected signal as observed by the
OWFA are presented in Marthi et al (2017, in preparation) based on
a simulation package (described in \citet{marthi16}). These
calculations include the HI signal, based on simulations of the expected
cosmological distribution of HI. The HI simulations themselves are
presented in \citet{chatterjee16}. More details regarding the predicted
outcomes of observations of HI with OWFA can be found in
\citet{ali14,bharadwaj15}, as well as in \citet{sarkar16}. \citet{ghelot16}
present a similar study of the HI signal expected to be seen by OWFA and also
compare OWFA with other instruments operating in this frequency range
as far as suitability for detection of HI at z$ \sim 3.3$ is concerned.

\subsection{Studies of the Inner Heliosphere}

Interplanetary scintillation (IPS) measurements are extremely useful to 
constrain the physical properties of the solar wind in the entire inner 
heliosphere (e.g., \cite{mano1993SoPh, mano2001ApJ}). For example, the 
legacy analog system of the ORT allows one to monitor IPS of about 
600--1000 radio sources per day. Such observations have played a crucial 
role in studies of 3-D evolution of coronal mass ejections (CMEs),
development of the associated interplanetary shocks as well
as the acceleration of particles in the Sun-Earth space
(e.g., \cite{mano2010SoPh, manoagalya2011advgeo}). However, the Ooty 3-D 
reconstruction of both solar wind speed and density obtained from 
such monitoring has been limited to a latitude and longitude resolution 
of $\sim$10$^\circ \times 10^\circ$ with about 10 {\rsun} increment in 
heliocentric distance \cite{mano2010SoPh}. This restriction arises
from the number of sources it is possible to observe in a given time.
It also takes about half a day of observation to provide a 6-hour increment
in 3-D views \cite{mano2010SoPh, manopankajwahab2011SoPh}.  In the case
of OWFA,  simulations show that the increase in sensitivity as well
as the considerably more powerful beamforming capabilities (as compared
to the legacy system) would allow for a $\sim 5$ fold increase in the
number of sources that could be monitored in a given time (\citet{manoharan16}).
This allows for the imaging of CMEs and solar wind structures with
much improved temporal and spatial  resolutions. This in turn would
provide  a better understanding of the physical processes associated
with the  evolution of CMEs (for e.g. the associated shock, sheath and
their influence on the ambient solar wind) and other solar wind transients.

\subsection{Transient Studies}

The good sensitivity and very large field of view of the OWFA instrument make
it an ideal instrument in blind searches for rare events such as radio
transients. Interest in radio studies of transient events has been steadily 
increasing in the recent past, and forms one of the key science goals of
many of the next generation telescopes including the SKA \citep{fender15}.
In particular the discovery of a new class of radio bursts with millisecond
duration and dispersion measures consistent with an origin in cosmologically
distant objects has led to renewed interest in blind searches for short duration
bursts \citep{lorimer07,thornton13}. Although most of these so called Fast Radio
Bursts (FRBs) have been detected at frequencies $\sim 1.4$~GHz, there has also
been a detection at lower frequencies ($\sim 700-900$~MHz) \citep{kiyoshi15}.
While the DMs  associated with FRBs are consistent with an extra-galactic
origin, the exact source of these bursts is as of yet unclear and there are
models in which the sources are local (see e.g. \citep{loeb14}).  Detailed
calculations of the expected FRB rate using the OWFA (assuming that the
sources are at cosmological distances) show that the detection rate could
be as high as several per day, depending on the assumed spectral index
\citep{bera16}. Clearly transient searches with the OWFA would be of high
scientific interest. More detailed calculations of the expected rate for
different assumed scattering models and different observing strategies with
the OWFA are presented in \citet{bhattacharyya16}.

\section{Summary}
\label{sec:sum}

This paper describes an ongoing upgrade to the legacy Ooty Radio Telescope
(ORT) to convert it into a wide field of view interferometer, called
the Ooty Wide Field Array (OWFA). The key elements of this upgrade are
briefly described, more details can be found in other papers in this
issue. There are three main science drivers to this upgrade, viz.
(1) Observations of large scale distribution of HI in the post-reionisation
era (2) studies of the propagation of plasma irregularities through
the inner heliosphere and (3) blind surveys for transient sources. The
key points of these three science cases have been discussed, interested
readers can find more details in the other papers in this issue.

\section*{Acknowledgment}

We are grateful to the staff at the Radio Astronomy Centre (RAC) Ooty,
whose help formed a critical component of this project. We also acknowledge
the assistance from Peeyush Prasad and T. C. Pawan during the realisation
of this project.

\bibliography{OWFAOverviewV3}{}
\bibliographystyle{plainnat}
\end{document}